\documentclass[aps,prl,superscriptaddress,reprint]{revtex4-1}

\usepackage{amsmath,amsfonts,amssymb}
\usepackage{graphicx}
\usepackage{hyperref}

\newcommand{\Var}{\mathrm{Var}}
\newcommand{\Cov}{\mathrm{Cov}}

\begin{document}

\title{State preparation and tomography of a nanomechanical \\resonator with fast light pulses}
\author{Juha T. Muhonen}
\email{juha.t.muhonen@jyu.fi}
\affiliation{Center for Nanophotonics, AMOLF, Science Park 104, 1098 XG, Amsterdam, The Netherlands}
\affiliation{Department of Physics and Nanoscience Center, University of Jyv{\"a}skyl{\"a}, P.O. Box 35, FI-40014 University of Jyv{\"a}skyl{\"a}, Finland}

\author{Giada R. La Gala}
\affiliation{Center for Nanophotonics, AMOLF, Science Park 104, 1098 XG, Amsterdam, The Netherlands}

\author{Rick Leijssen}
\affiliation{Center for Nanophotonics, AMOLF, Science Park 104, 1098 XG, Amsterdam, The Netherlands}

\author{Ewold Verhagen}
\email{verhagen@amolf.nl}
\affiliation{Center for Nanophotonics, AMOLF, Science Park 104, 1098 XG, Amsterdam, The Netherlands}

\begin{abstract}
Pulsed optomechanical measurements enable squeezing, non-classical state creation and backaction-free sensing. We demonstrate pulsed measurement of a cryogenic nanomechanical resonator with record precision close to the quantum regime. We use these to prepare thermally squeezed and purified conditional mechanical states, and to perform full state tomography. These demonstrations exploit large photon-phonon coupling in a nanophotonic cavity to reach a single-pulse imprecision of 9 times the mechanical zero-point amplitude $x_\mathrm{zpf}$. We study the effect of other mechanical modes which limit the conditional state width to $58x_\mathrm{zpf}$, and show how decoherence causes the state to grow in time.
\end{abstract}

\maketitle

Measurement and control of mechanical motion at the quantum level is of wide interest because of the quantum technologies it would enable and the possibility to probe decoherence in massive quantum systems. Cavity optomechanical demonstrations \cite{Aspelmeyer2014} of quantum control over mechanical resonators included ground state cooling \cite{OConnell2010,Chan2011,Teufel2011}, quantum squeezing \cite{Wollman2015,Pirkkalainen2015,Lecocq2015}, entanglement \cite{Riedinger2017,Korppi2017}, and exchanging individual quanta between the mechanical oscillator and qubits \cite{OConnell2010,Chu2017}. 

Measurement and control are intimately linked. At a basic level, any measurement process is a competition between information gain (measurement rate), decoherence processes, and noise added to the system by the measurement (backaction). If the measurement rate overcomes those detrimental effects, control over the system can be achieved, allowing for mechanical oscillators for example feedback cooling to the ground state \cite{Cohadon1999,Wilson2015,Rossi2018}. A continuous measurement of a mechanical resonator's displacement is subject to the so-called standard quantum limit (SQL). The SQL is a manifestation of Heisenberg's uncertainty principle as these measurements simultaneously observe the two non-commuting motional quadratures, hence giving a lower limit, equal to the zero-point fluctuation amplitude $x_\mathrm{zpf}$, to the noise added by the measurement \cite{Murch2008,Purdy2013}. Methods to evade this backaction limit by moving away from simple continuous displacement measurements were demonstrated in various quantum systems \cite{Vasilakis2015,Korppi2016, Moeller2017,Kampel2017,Sudhir2017,Mason2018}, including in sideband-resolved opto- or electromechanical cavities probed by two-tone fields \cite{Hertzberg2009,Suh2014,Shomroni2018}, where the measurement is only sensitive to one motional quadrature.

Another method for backaction-evading measurements that has been put forward is that of pulsed measurements, where mechanical motion can be neglected during a short interaction \cite{Braginsky1978,Vanner2011}. A single `snapshot' measurement of a harmonic oscillator's position measures one of the quadratures with potentially unlimited precision, as all backaction is introduced to the orthogonal quadrature with no effect to the future evolution of the measured quadrature. With sufficient precision the mechanical oscillator is then prepared to a squeezed state, conditioned on the measurement. Importantly, the ability to probe either quadrature precisely allows full quantum state tomography \cite{Vanner2014}. Combined with nonlinear displacement measurement or non-Gaussian states of light, pulsed interactions can also induce other nonclassical states \cite{Vanner2011a,Hoff2016}. Moreover, proposals recently suggested to exploit pulsed measurement for swap operations between mechanics and light \cite{Bennett2016} and creation of macroscopic superpositions \cite{Hoff2016}.

Despite these extensive theoretical proposals, so far only a single proof-of-principle experimental demonstration of pulsed optomechanical measurements has been reported \cite{Vanner2013}, at elevated temperature and without cavity enhancement. In order to reach quantum-level accuracy with a single pulsed measurement one needs to fulfill the challenging requirement $8\eta\sqrt{N_P}g_0/\kappa\gtrsim1$, where $g_0$ is the cavity frequency shift for a displacement $x_\textrm{zpf}$ (the photon-phonon coupling rate), $\eta$ the coupling efficiency of light to the cavity, and $N_P$ the number of photons in the pulse \cite{Vanner2011}, while at the same time offering sufficient cavity linewidth $\kappa$ to accommodate a pulse of duration $\tau_P\ll2\pi/\omega_m$, where $\omega_m$ is the mechanical oscillation frequency.

In this work, we address these challenges using a cavity optomechanical system based on a sliced photonic crystal nanobeam, allowing large photon-phonon coupling rates \cite{Leijssen2015,Leijssen2017}. We demonstrate pulsed optomechanical measurements close to the quantum regime, achieving a record-low shot-noise limited single-pulse measurement imprecision of $9x_\mathrm{zpf}$, constrained mainly by the optical detection efficiency. We prepare both thermally squeezed and purified (cooled) conditional mechanical states, and perform full state tomography on these states. We study how additional mechanical modes affect the conditional state, limiting its width in this experiment to $58x_\mathrm{zpf}$. We also demonstrate how thermal dephasing can be tracked by recording the evolution of the state at longer timescales. In addition, we show how post-selection allows maximizing measurement sensitivity even though the large optomechanical interaction strength pushes our system deep into the regime of nonlinear optomechanical interaction where sensitivity is reduced \cite{Leijssen2017}.

\begin{figure}
\includegraphics[width=\columnwidth]{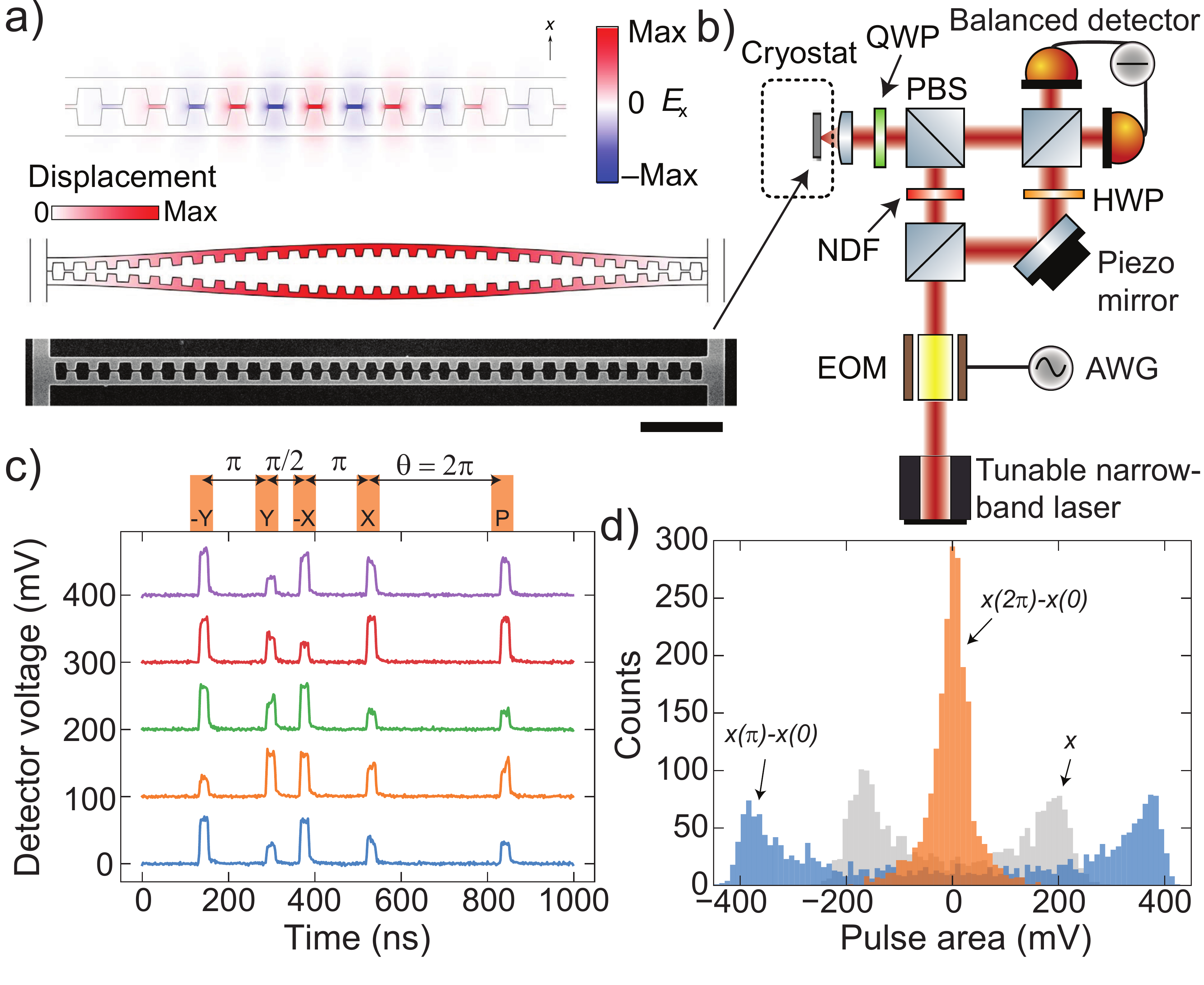}
\caption{\textbf{(a)} (Bottom) Electron microscope image of the free-standing, 250~nm thick sliced silicon nanobeam. Scale bar is 2 $\mu$m. (Middle) Simulated displacement profile of the flexural vibrations of the two beam halves. (Top) Simulated transverse electric field of the optical cavity mode. 
\textbf{(b)} Schematic diagram of the employed balanced homodyne interferometer set-up. Pulses are created by an electro-optical-modulator (EOM) controlled by an arbritrary waveform generator (AWG). Quarter-wave-plate (QWP), half-wave plate (HWP), neutral density filter (NDF) and polarising beamsplitter (PBS) are also shown.
\textbf{(c)} Example measurement traces, and schematic of the used pulsing sequence (separations $\omega t$ indicated). The different measurement traces are offset for clarity.
\textbf{(d)} Examples of extracted histograms: thermal state [random sampling, mean subtracted] (grey), difference of two pulses a half-period apart (blue) and difference of two pulses a full period apart (orange).
\label{fig:one}}
\end{figure}

Figure~\ref{fig:one}(a) shows a diagram of the sliced silicon nanobeam. The device, presented in Ref.~\onlinecite{Leijssen2017}, hosts a photonic crystal nanocavity mode whose resonance frequency (204 THz) depends very sensitively on the flexural movement of the two beam halves. These move roughly independent of each other with two mechanical mode frequencies around $\omega_m/(2\pi)\approx3$~MHz, separated by $\sim$120 kHz. The photon-phonon coupling rate $g_0/(2\pi) \approx 25$ MHz is approximately equal for both modes. The optical cavity linewidth $\kappa/(2\pi)\approx20.4$ GHz enables practically instantaneous measurements of mechanical position while still achieving $g_0/\kappa>10^{-3}$. All reported measurements are performed on the same sample at a temperature of 3.2 K. The exact mechanical frequencies and damping rates drift with time for reasons not fully understood, and are mentioned in the captions for each dataset.

The sample is incorporated in a homodyne interferometer (Fig.~\ref{fig:one}(b)). An electro-optic modulator driven by a waveform generator produces optical pulses using light from a continous-wave tunable narrowband diode laser, which are sent into both interferometer arms. Light is focused on the sample through an $\mathrm{NA}\approx0.55$~lens, coupling to the nanocavity with efficiency $\eta=\sqrt{\eta_\mathrm{in}\eta_\mathrm{out}}\approx0.01$ \cite{Leijssen2017}. Here $\eta_\mathrm{in, out}$ are the efficiencies with which light is coupled from the incident laser beam to the cavity and from the cavity to the detectors, respectively. Each incident pulse carries $\sim2\times10^6$ photons in a duration of $\tau_P=20$~ns, such that the estimate maximum number of simultaneous intracavity photons is $\sim60$. The same lens collects emitted cavity radiation, whose phase quadrature is measured by recording the output of a balanced detector after interference with the local oscillator pulse. The resultant detector voltage thus reflects optical phase and mechanical displacement $x$.

Figure~\ref{fig:one}(c) shows examples of recorded pulse traces. Between traces we wait for a time (30~ms) larger than the mechanical damping time $2\pi/\Gamma$. It can be directly recognized that the recorded pulse heights are correlated when they are separated by a full oscillation period (two last pulses) whereas the pulses separated by half a period (e.g., the third and fourth pulse) are anti-correlated (around a non-zero offset voltage). Figure~\ref{fig:one}(d) depicts histograms of the difference of the recorded pulses (integrating the voltage over the pulse duration), demonstrating this correlation and anti-correlation behaviour. This is direct indication that the thermal nanobeam motion is imprinted on the detected pulses.

\begin{figure}
\includegraphics[width=\columnwidth]{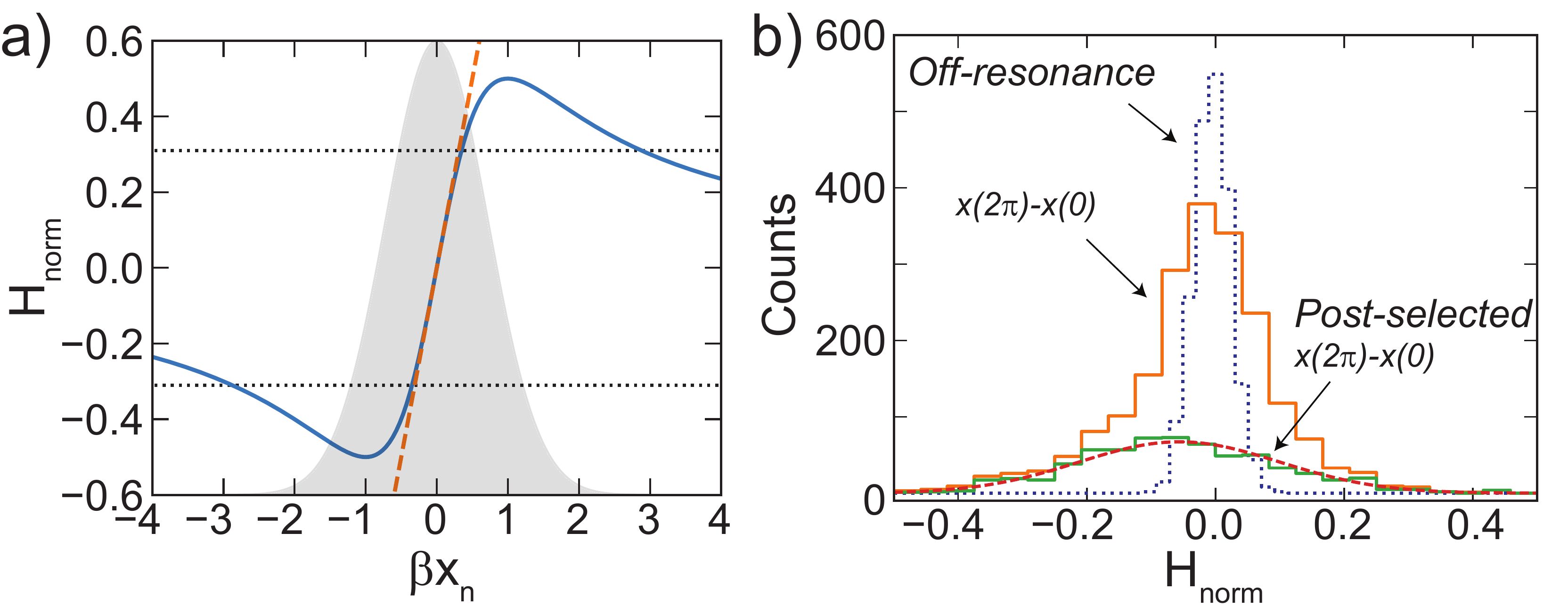}
\caption{\textbf{(a)} Full transduction function (Eq.~\ref{eq:homodyne}, solid line), showing the linear approximation (dashed line) and the post-selection threshold (dotted lines). Gray area depicts a Gaussian with variance $\beta^2\sigma_\mathrm{th}^2$. 
\textbf{(b)} The histogram of the difference of two pulses a full period apart (as in Fig.~\ref{fig:one}(d)), before and after post-selection (solid lines). After post-selection we recover a Gaussian shape (dashed line shows fit). Also shown is a reference measurement off-resonant with the optical cavity (dotted line), from which the measurement noise floor is extracted.
\label{fig:two}}
\end{figure}

To understand the histogram shapes, we need to consider the nonlinearity of the transduction between mechanical position and optical phase. At the phase-sensitive operation point of the homodyne interferometer and on resonance with the optical cavity the balanced detector output is~\cite{supp}
\begin{equation}
H = 4|a_\mathrm{in}||a_\mathrm{lo}|\eta\frac{\beta x_n}{\beta^2x_n^2+1},
\label{eq:homodyne}
\end{equation}
with $\beta=2g_0/\kappa$, $x_n=x/x_\mathrm{zpf}$ the normalized displacement and $a_\mathrm{in}$ and $a_\mathrm{lo}$ the optical field amplitudes towards the sample and in the local oscillator, respectively. This homodyne signal (Fig.~\ref{fig:two}(a)) depends linearly on displacement only when $\beta x_n\ll1$. Outside this regime, the relationship between signal and displacement is nonlinear and even multivalued. Therefore, our value of $\beta\approx2.5\times 10^{-3}$ causes the thermal Gaussian displacement due to both modes with width (square-root of variance) $\sigma_\mathrm{th}\approx290x_\mathrm{zpf}$ to be distorted into the double-peaked probability histogram in Fig.~\ref{fig:one}(d). To calibrate the homodyne signal we fit this thermal histogram to an analytical model using the known sample temperature~\cite{supp}. The fit allows converting the measurement signal to normalized homodyne signal $H_\textrm{norm}=H/4|a_\mathrm{s}||a_\mathrm{lo}|\eta$.

The pulsing sequence in Fig.~\ref{fig:one}(c) is used for conditional state preparation and tomography. The pulses are separated by an angle $\theta\equiv\omega_m t$, where $\omega_m$ is the frequency of the mechanical mode of interest. (We address the impact of multiple modes below.) The mechanical resonator motion can be written as $x(\theta)=X\cos(\theta)+Y\sin(\theta)$, using the quadrature amplitudes $X,Y$ which vary slowly within pulse trains but are randomized between pulse trains with zero mean and variance $\Var(X)=\Var(Y) = 2n_\mathrm{th}x_\mathrm{zpf}^2$, where $n_\mathrm{th}=k_BT/(\hbar\omega_m)$ is the number of thermal phonons. The four first pulses of the sequence measure the instantaneous value of the two quadrature amplitudes (state preparation) and the last pulse quantifies the difference between the expected mechanical position and the actual position (tomography). By varying the waiting time between the state preparation and the tomography pulse, we can map out this difference in all quadrature angles and perform full state tomography of the conditional state. Note that we measure both $-X$ and $X$ using two pulses to cancel measurement offsets caused by low-frequency drifts~\cite{supp}.

To maximize sensitivity and allow single-valued estimation, we post-select the data so that the measured value for the quadrature amplitude of interest falls in the linear transduction regime. The dotted lines in Fig.~\ref{fig:two}(a) show the chosen post-selection thresholds. Figure~\ref{fig:two}(b) shows the effect of post-selection on the histogram of the difference of two pulses separated by $\theta = 2\pi$. The original histogram had non-Gaussian shape with variance dominated by shot-noise due to a large contribution from parts of the transduction function with reduced sensitivity at $|\beta x_n|\approx1$. The post-selected histogram has a Gaussian shape with larger variance. Hence, the post-selection protocol allowed recovering the linear operating regime.  

To understand the principles of state preparation, consider a measurement at $t=0$, measuring $X$. With just this measurement, the prediction for the oscillator's position at later times is $\cos(\theta)X$, differing from the actual position by $x(\theta)-\cos(\theta)X=\sin(\theta)Y$ (with variance $\sin^2(\theta)\Var(Y)=2n_\mathrm{th}x_\mathrm{zpf}^2\sin^2(\theta)$ for a thermal state). This notably goes to zero at $\theta=\{\pi,2\pi\}$, demonstrating that the knowledge of one quadrature allows ideally predicting the mechanical position exactly every half-period. In a phase-space defined by the two quadrature amplitudes this is a squeezed state.
Combining two ideal measurements (a quarter period apart) then allows to measure both quadratures and fully predict the oscillator's evolution. In other words, we have then prepared a pure state with no classical uncertainty in $x$. This pure state will then decay towards the thermal equilibrium distribution with the time constant $2\pi/\Gamma$.

In the above idealized case, the pulsed measurements are assumed to be infinitely accurate and backaction-free. In practice, finite measurement imprecision leads to a Gaussian probability distribution for the measured quadrature amplitude. In addition, unavoidably, any measurement disturbs the other quadrature. These intuitions are formalized with a measurement operator given by Caves and Milburn for a free particle \cite{Caves1987} and Vanner \textit{et al.} for harmonic motion \cite{Vanner2011} 
\begin{equation}
\hat{M} = \frac{1}{\sqrt[4]{\pi}}\exp\left(i\Omega \hat{\mathcal{X}} - \frac{(\hat{\mathcal{X}}-\mathcal{M})^2}{2/\chi^2}\right),
\end{equation}
where $\mathcal{M}= x_n^\textrm{meas}/\sqrt{2}$ and $x_n^\textrm{meas}$ is the dimensionless measurement result (normalized by $x_\mathrm{zpf}$) and $\hat{\mathcal{X}}=(\hat{b}^\dag+\hat{b})/\sqrt{2} = \hat{x}/(\sqrt{2}x_\mathrm{zpf})$ is the quadrature operator with $\hat{b}$ the phonon annihilation operator.
The parameter $\chi = 8\sqrt{\eta_\mathrm{in}\eta_\mathrm{out}N_P}g_0/\kappa$ \cite{Vanner2011,Hoff2016} characterizes the pulsed measurement strength and hence the conditional state variance. Physical insight into $\chi$ is provided by noting that the signal-to-noise-ratio of a shot-noise limited measurement is $\chi x_n$ \cite{supp}, meaning that with $\chi=1$ one measures a displacement of $x_\mathrm{zpf}$ with unity signal-to-noise ratio when comparing two pulses.

Performing a measurement (transforming the state with $\rho_f=\hat{M}\rho_i\hat{M}^\dag$) transforms one quadrature of an arbitrary initial state ($\rho_i$) into a Gaussian with width $\sigma_m=x_\mathrm{zpf}/\chi$ and mean given by the random measurement result (which follows statistics determined by $\rho_i$), while adding $\Omega\sqrt{2}x_\mathrm{zpf}$ to the other quadrature. That quadrature will also gain uncertainty  $\sigma_\mathrm{ba} = \sqrt{\eta_\mathrm{in}/\eta_\mathrm{out}}\chi x_\mathrm{zpf}$, hence forcing for a state prepared by two sequential measurements quarter-period apart $\sqrt{\sigma_m(\sigma_m+\sigma_\mathrm{ba})} \geq x_\mathrm{zpf}$.
In Fig.~\ref{fig:two}(b) we show a histogram measured away from optical resonance to determine the noise floor of the measurement. From this we can extract a single-pulse measurement imprecision of $\sigma_m\approx9x_\mathrm{zpf}$ \cite{supp}, corresponding to $\chi\approx0.11$, some three orders of magnitude higher than previous state-of-the-art \cite{Vanner2013}. This value is consistent with measured sample parameters including $\eta_\mathrm{in} \approx 1.3 \%$ and for $\eta_\mathrm{out}\approx0.35\eta_\mathrm{in}$, close to our previous independent estimation \cite{Leijssen2017}.

\begin{figure}
\includegraphics[width=\columnwidth]{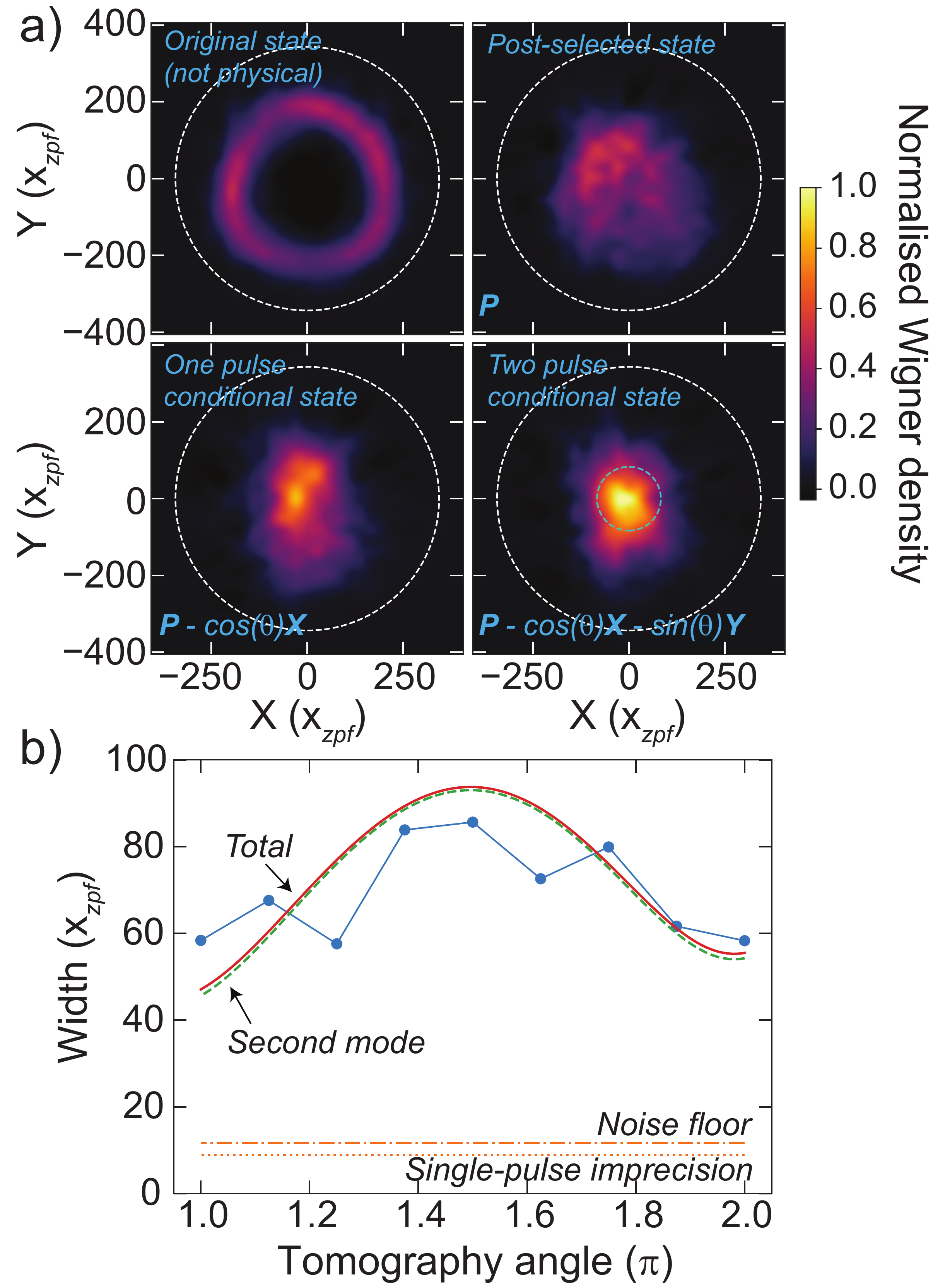}
\caption{\textbf{(a)} Reconstructed Wigner densities, for non-post-selected data (top left), post-selected data (top right), one-pulse conditional data (bottom left) and two-pulse conditional data (bottom right).
A single normalization is applied on all panels. 
White dashed line shows the calculated full-width-half-maximum (FWHM) of the original thermal state at 3.2~K, and green dashed line the measured FWHM of the conditional state. Each panel is reconstructed from nine different measurement angles $\theta$ each combining 2000 acquired traces. 
\textbf{(b)} Width of different mechanical marginals for the two-pulse conditional data.
Also shown are the noise floor of the measurement (dash-dotted line), the expected contribution from the second mechanical mode (dashed line) and the squared sum of these (solid line). Note that the noise floor differs slightly from the single-pulse imprecision (dotted line) as multiple pulses are used in the measurement and shot noise of the tomography pulse is not subtracted~\cite{supp}. The two mechanical frequencies were $\omega=2\pi\times3.1081$ MHz and $r\omega=2\pi\times3.2280$ MHz.
\label{fig:three}} 
\end{figure}

Figure \ref{fig:three} presents experimental results of conditional state preparation and tomography. Scanning the delay between preparation pulses and the tomography pulse (i.e., varying $\theta$) allows mapping the mechanical marginals and reconstructing the phase-space Wigner function of the conditional state via an inverse Radon transform from the histograms \cite{Vanner2011}. As we subtract the measured (random) values for the quadrature amplitudes, the plots depict the conditional state shifted to origin.
The upper panels of Fig.~\ref{fig:three}(a) show this Radon transform for only the tomography pulse, without (left) and with (right) post-selection. These should be circularly symmetric and without the nonlinearity would simply depict the thermal Gaussian distribution. The post-selected data now depicts a small central part of that distribution. The lower left panel shows the thermally squeezed state prepared with one quadrature conditionalization. The state is extracted with $P-\cos(\theta)X$, with $P$ the result of the tomography pulse and $X$ the measured quadrature amplitude from the previous pulses. 
Finally, in the lower right panel of Fig.~\ref{fig:three}(a), we plot the Wigner function for the state conditionalized in both quadratures $P-\cos(\theta)X-\sin(\theta)Y$, which would have the same area as the ground state if the measurements would be ideal. In our case the state has an average width that corresponds to a one mode thermal state at a temperature of 380 mK, reflecting purification from the original temperature of 3.2 K.

The width of the state conditionalized on both quadratures is plotted in Fig.~\ref{fig:three}(b) as a function of the tomography angle. In the measured data the minimum width reached is $58x_\mathrm{zpf}$, significantly exceeding the shot noise floor, and is maximized at $\theta=3\pi/2$. These features are explained through the existence of a second mechanical mode. As in this device the two mechanical modes couple equally strongly to the cavity, the contribution of the second mechanical mode is captured with $x(\theta)=X_1\cos(\theta)+Y_1\sin(\theta)+X_2\cos(r\theta)+Y_2\sin(r\theta)$, with $r$ the ratio of the mechanical frequencies and subscripts refer to the two modes. The resulting uncertainty caused by the second mode \cite{supp} as a function of $\theta$ is plotted in Fig.~\ref{fig:three}(b), matching the data well. There are no fitting parameters here as frequency and temperature of both modes are known. The non-monotonic shape is caused by measuring the $Y$ quadrature before the $X$ quadrature, causing it to have a larger contribution as it has more time to evolve out-of-sync with the mode of interest~\cite{supp}. 

\begin{figure}
\includegraphics[width=\columnwidth]{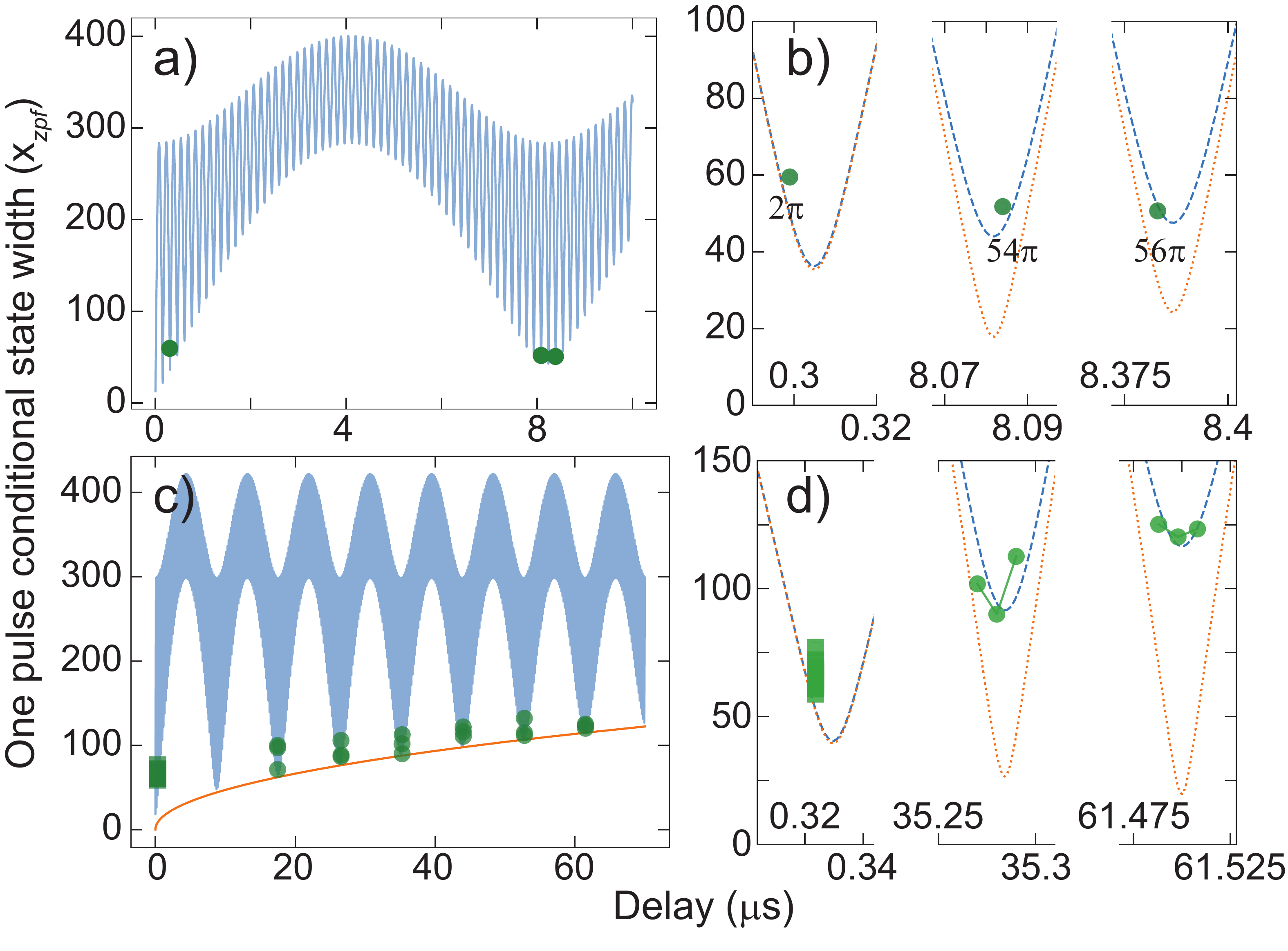}
\caption{
\textbf{(a-b)} Common mode measurement, showing that smaller conditional state width is achieved at $\theta=54\pi,56\pi$ than at $\theta=2\pi$ due to minimization of the second mode contribution at these times. Panel (b) shows close-ups of (a). Blue line shows the expected behaviour with fitted $\Gamma/(2\pi)=400$~Hz, yellow dashed line [only shown in (b)] without any decoherence. Circles are measured data. Frequencies $\omega=2\pi\times3.340$~MHz and $r\omega=2\pi\times3.218$~MHz.
\textbf{(c-d)} Thermal decoherence measurement demonstrating the growing conditional state width. Panel (d) shows close-ups of (c). Yellow line in (c) is the envelope function $\sqrt{[1-\exp(-t\Gamma/2)]8n_\mathrm{th}}x_\mathrm{zpf}$, and the blue line the full expected curve, both with fitted $\Gamma/(2\pi)=400$~Hz, yellow dashed line in (d) without decoherence. Each pulse sequence here has two tomography pulses, one at $\theta=2\pi$ (shown as squares) and other at variable distance (circles). Also shown is data at $\pm 10$~ns around $\theta=2n\pi$ points. Frequencies $\omega=2\pi\times3.090$~MHz and $r\omega=2\pi\times 2.976$~MHz. All datasets contained 1000 samples before post-selection.
\label{fig:four}}
\end{figure}

Assuming that $r\approx1$ and that the two modes have equal $n_\mathrm{th}$ and $x_\mathrm{zpf}$, the expected contribution to the conditional state width from the second mode can be approximated as $\sqrt{4n_\mathrm{th}\left[1-\cos(r\theta)\cos(\theta)-\sin(r\theta)\sin(\theta)\right]}x_\mathrm{zpf}$ \cite{supp}. This is expected to vanish when $\cos(\theta)\cos(r\theta)=1$ and $\sin(\theta)\sin(r\theta)=0$, or vice versa. Although this cannot be fulfilled exactly unless $r$ is a rational number, it is approximated when $\theta=2n\pi$, where $n\approx\omega_m/(r\omega_m-\omega_m)$. In Fig.~\ref{fig:four}(a-b) we compare the one-pulse conditional state width after one mechanical period ($\theta=2\pi$) and where this condition is fulfilled ($n\approx54,56$).
A slightly lower conditional state width is measured at $\theta=54\pi,56\pi$ than at $\theta=2\pi$.

The measured conditional state widths at $\theta=54\pi,56\pi$ differ strongly from expectation based on the formula above. This is because we neglected thermal decoherence, which will cause the minimum conditional state width to increase in time as $\sqrt{8n_\mathrm{th}[1-\exp\left(-t\Gamma/2 \right)]}x_\mathrm{zpf}$, in the case of two mechanical modes, assuming the modes have identical $n_\mathrm{th}$, $x_\mathrm{zpf}$ and $\Gamma$ \cite{supp}. Figure~\ref{fig:four}(c-d) shows measurements around the times where the second mode contribution should vanish on a longer timescale, tracking the loss of coherence due to thermalization quantitatively in time-domain.

These results show that nano-optomechanical systems can bring quantum-level mechanical measurements with single nanosecond pulses within reach. Notably, achieving $\eta\gtrsim8\%$ without changing any other parameters would bring the uncertainty in one quadrature below $x_\mathrm{zpf}$, allowing squeezed state preparation and observing quantum backaction \cite{Purdy2013}. 
Preparing a pure state of a single resonator would require reducing the second mode contribution, through (opto)mechanical design (coupling the two beam halves more strongly to create a single optically bright mode), or by further cooling (cryogenic or feedback). Alternatively, one could exploit the fact that with quantum-level precision, a single pulse would entangle the quadratures of the two mechanical modes, providing a new path to explore many-mode quantum optomechanics in the time domain.

Indeed, our experiments demonstrate how pulsed measurements yield interesting possibilities for measurement and control of mechanical motion, complementing the conventional frequency domain analyses. This ``time-domain optomechanics'' may give rise to new protocols for quantum sensing that exploit the fast backaction-free determination of a mechanical quadrature, as well as new paradigms to create quantum states of motion and mechanical entanglement.

\textbf{Acknowledgements.}
The authors thank Hugo Doeleman for critical reading of the manuscript. This work is part of the research programme of the Netherlands Organisation for Scientific Research (NWO), and supported by the European Union's Horizon 2020 research and innovation programme under grant agreement No 732894
(FET Proactive HOT). E.V. gratefully acknowledges an NWO-Vidi grant for financial support. J.T.M. thankfully acknowledges funding from the European Union's Horizon 2020 research and innovation programme under the Marie Sklodowska-Curie grant agreement No 707364.

\clearpage
\onecolumngrid

\renewcommand{\thefigure}{S\arabic{figure}}
\renewcommand{\thetable}{S\arabic{table}}
\renewcommand{\theequation}{S\arabic{equation}}

\setcounter{figure}{0}
\setcounter{equation}{0}

\section*{Supplementary Information}

\begin{figure}[h]
\includegraphics[width=0.45\textwidth]{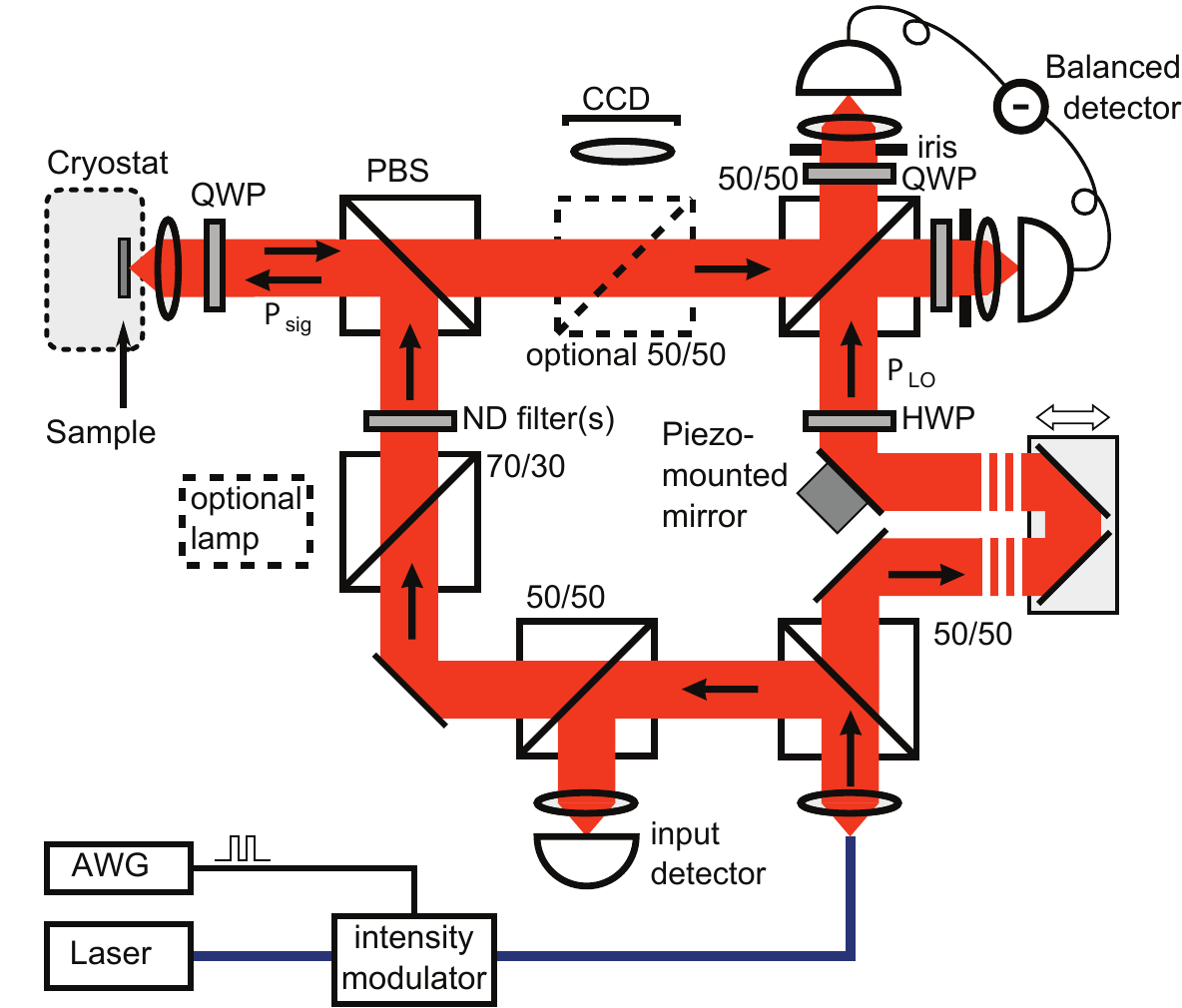}
\caption{\textbf{Detailed measurement setup.} The lamp and the CCD are used to align the laser beam with our sample using a moving lens outside the cryostat window. The extra quarter wave plates (QWP) in front of the detector are for preventing detector reflections of the local oscillator from reaching the sample as they will be reflected by the polarising beam splitter. Irises are used to balance the two detectors. Piezo mounted mirror is used to adjust the local oscillator phase with respect to the signal phase. $P_
\mathrm{lo}$ and $P_\mathrm{sig}$ refer to the power in the local oscillator arm and signal arm, respectively.
\label{fig:supp_setup}} 
\end{figure}

\subsection{Details of the pulsing sequences and single-pulse imprecision extraction}

All pulse trains we apply are separated from each other by a time that is large compared to the mechanical damping time in order to 'reset' the mechanical state ($\sim$30 ms). For reasons not fully understood (but most likely deriving from the fact that our homodyne phase is not actively locked during measurements and/or technical laser noise) our measured voltages have an offset that is constant inside a pulse train but varies between different pulse trains. In order to cancel out this offset we always take two pulses per quadrature for the state preparation, as well as subtract this offset from the tomography pulse.

The pulse train for two pulse conditionalization is shown in Fig.~\ref{fig:one}(c). Assuming the signal is given by $s(\omega t) = X\cos(\omega t) + Y\sin(\omega t) + V_\mathrm{off} + \delta$, where $V_\mathrm{off}$ is the offset and $\delta$ random noise, and setting $t=0$ at the final pulse before the tomography pulse, the four preparation pulses read
\begin{eqnarray}
P_0 &=& s(-5\pi/2) = -Y+V_\mathrm{off}+\delta_0 \\
P_1 &=& s(-3\pi/2) = Y+V_\mathrm{off}+\delta_1 \\
P_2 &=& s(-\pi) = -X+V_\mathrm{off}+\delta_2 \\
P_3 &=& s(0) = X+V_\mathrm{off}+\delta_3.
\end{eqnarray}

Hence, to extract the quadrature amplitudes we use $Y=(P_1-P_0)/2$ and $X=(P_3-P_2)/2$. Then the full two pulse conditional state reads
\begin{equation}
s_\mathrm{cond}(\omega t) = s(\omega t)-\frac{P_3-P_2}{2}\cos(\omega t)-\frac{P_1-P_0}{2}\sin(\omega t)-\frac{P_0+P_1+P_2+P_3}{4},
\label{eq:pulseSequence}
\end{equation}
where the last term is used to cancel the offset in the final tomography pulse $s(\omega t)$. Note that replacing that term with either $(P_0+P_1)/2$ or $(P_3+P_2)/2$ cancels out some of the noise terms in one quadrature at the expense of increasing the noise in the other quadrature. For the one-pulse conditional state in Fig.~\ref{fig:three} we use the same formula without the sine term, and for the non-conditional state we also remove the cosine term (but keep the offset subtraction term).

The noise floor presented in the main text Fig.\ref{fig:three}(b) was extracted by running the same pulse sequence and analysis for measurement data gathered with laser tuned off-resonance with the optical cavity. The resulting conditional state histogram with $\theta=2\pi$ is presented in main Fig.~\ref{fig:two}(b), and has the width $11.67x_\mathrm{zpf}$, when converted to displacement. As the off-resonance data will presumably have no signal ($X,Y$=0) we can write
\begin{eqnarray}
s^\mathrm{noise}_\mathrm{cond}(2\pi) &=& s(2\pi)-\frac{P_3-P_2}{2}\cos(2\pi)-\frac{P_0+P_1+P_2+P_3}{4} \\
&=& V_\mathrm{off}+\delta_4 -\frac{V_\mathrm{off}+\delta_3-V_\mathrm{off}-\delta_2}{2}-\frac{4V_\mathrm{off}+\delta_0+\delta_1+\delta_2+\delta_3}{4} \\
&=& \delta_4 -\frac{\delta_0+\delta_1}{4}-\frac{3\delta_3}{4}+\frac{\delta_2}{4}.
\end{eqnarray}
Hence, we have for the variance (we assume all noise terms are independent and hence have no mutual covariance)
\begin{eqnarray}
\Var\left[s^\mathrm{noise}_\mathrm{cond}(2\pi)\right] &=& \Var(\delta_4) +\frac{\Var(\delta_0)+\Var(\delta_1)+\Var(\delta_2)+9\Var(\delta_3)}{16} \\
&=& (1+\frac{3}{4})\Var(\delta_\mathrm{sn}),
\end{eqnarray}
where for the last line we have assumed that the variance of noise in each pulse is the same and marked it $\Var(\delta_\mathrm{sn})$. The noise floor variance is the sum of the imprecision in the measurement of $X$ [$\Var(\delta_\mathrm{sn})/2$ as two pulses are used], the imprecision in the tomography pulse [$\Var(\delta_\mathrm{sn})$] and the added variance due to the offset correction [$\Var(\delta_\mathrm{sn})/4$].
Hence, in order to extract the single-pulse imprecision $\sqrt{\Var(\delta_\mathrm{sn})}$ we need to divide the variance we extract by $7/4$, meaning we need to divide the width we extracted $11.67x_\mathrm{zpf}$ with square root of $7/4$, giving $8.8x_\mathrm{zpf}$.

We can make an easy check on this, as we can also extract the variance of the "non-conditional state" from the off-resonance data
 \begin{eqnarray}
s^\mathrm{noise}_\mathrm{non-cond}(2\pi) &=& s(\omega t)-\frac{P_0+P_1+P_2+P_3}{4} \\
&=& \delta_4 -\frac{\delta_0+\delta_1+\delta_2+\delta_3}{4},
\end{eqnarray}
from which using similar arguments as before we can extract $\Var\left[s^\mathrm{noise}_\mathrm{non-cond}(2\pi)\right] = (1+1/4)\Var(\delta_\mathrm{sn})$. From the data (not shown) we can extract a non-conditional width of $9.92x_\mathrm{zpf}$ which would again give a single pulse shot-noise of $8.8x_\mathrm{zpf}$.

Note that the actual imprecision in the measurement of $X$ (without the tomography pulse) is half the single-pulse imprecision, due to the fact that we use two pulses to measure it.

\subsection{Homodyne signal calibration}

As mentioned in the main text (and is derived below) the homodyne signal as a function of the normalized displacement can be written as 
\begin{equation}
H = A \frac{\beta x_n}{\beta^2x_n^2+1} \equiv A \frac{\Delta}{\Delta^2+1},
\end{equation}
where we have made explicit the relative change in optical cavity frequency due to the mechanical motion $\Delta = \beta x_n$ and used a constant $A$ to absorb all the normalization terms. Inverting this equation gives
\begin{equation}
\Delta^\pm = \frac{1\pm\sqrt{1-4H'^2}}{2H'},
\label{eq:delta}
\end{equation}
where we have defined $H'\equiv H/A$. As expected,this is multivalued and hence we cannot generally reliably deduce the mechanical position from a single measurement. This is only possible in the linearised regime $\beta^2 x_n^2 \ll 1$, where $H\approx A\beta x_n$. Hence, in the main manuscript we only use data where this approximation is valid (post-selection).

\begin{figure}[h]
\includegraphics[width=0.35\textwidth]{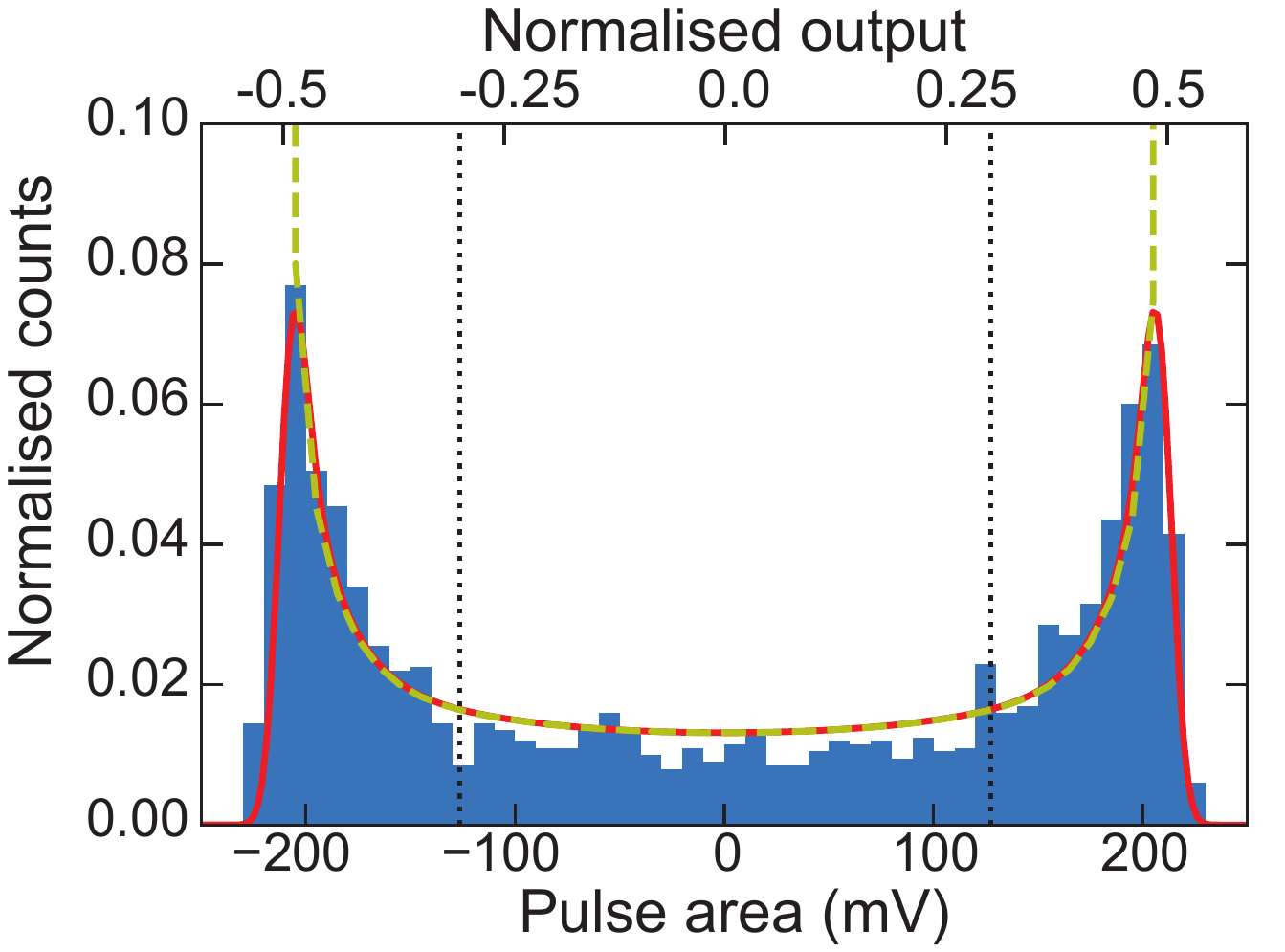}
\caption{Example of a thermal probability histogram of measurement outputs and fit to Eq.~\ref{eq:output} (dashed line). From this we can relate the measured voltages (bottom x-axis) to the normalized output ($H'$, top x-axis). Solid line shows a numerical convolution of the analytical solution (that diverges at $H'=\pm0.5$) and a Gaussian with the width given by our noisefloor. The thermal histogram points are extracted by taking the difference of two pulses half an oscillation period apart and dividing by two. This eliminates errors from a constant offset drift (see previous section). Time between individual measurements is roughly three times the mechanical damping time. Dotted lines show our post-selection thresholds at $H'=\pm0.31$.
\label{fig:supp_calibration}} 
\end{figure}

We can still, however, reliably predict ensemble histograms and this is used for calibrating the measurement signal. At thermal equilibrium $x_n$ will have a Gaussian probability density with variance $\Var(x_n) = (2k_BT)/(\hbar\omega_m)=2n_\mathrm{th}$, where $k_B$ is the Boltzmann constant, $T$ is temperature, $\hbar$ the reduced Planck constant and $\omega_m$ the mechanical oscillation frequency. For multiple independent mechanical modes, we need to sum the variances. The parameter $\beta$ is approximately equal for both modes. It follows that $\Var(\Delta) = \beta^2\Var(x_n) = 8(g_0/\kappa)^2 (n_\mathrm{th,1}+n_\mathrm{th,2})\equiv \sigma_{\Delta}^2$, where we have marked $n_\mathrm{th,i}$ as the mean thermal phonon number for mode $i$.
From probability calculus
\begin{equation}
\mathcal{P}(H') = \mathcal{P}(\Delta(H'))\left|\frac{d\Delta}{dH'}\right|=\frac{1}{\sqrt{2\pi\sigma_{\Delta}^2}}\exp\left(\frac{\Delta^2}{2\sigma_{\Delta}^2}\right)\left|\frac{d\Delta}{dH'}\right|,
\end{equation}
where we use $\mathcal{P}$ to mark the probability density. In order to deal with the multivaluedness we need to sum over both branches
 \begin{equation}
\mathcal{P}(H') = \frac{1}{\sqrt{2\pi\sigma_{\Delta}^2}}\exp\left(\frac{(\Delta^{+})^2}{2\sigma_{\Delta}^2}\right)\left|\frac{d\Delta^+}{dH'}\right| + \frac{1}{\sqrt{2\pi\sigma_{\Delta}^2}}\exp\left(\frac{(\Delta^{-})^2}{2\sigma_{\Delta}^2}\right)\left|\frac{d\Delta^-}{dH'}\right|.
\label{eq:output}
\end{equation}
Inserting Eq.~\ref{eq:delta} and its derivative
\begin{equation}
\left|\frac{d\Delta^\pm}{dH'}\right| = -\frac{1}{2H'}\left(1\mp\frac{1}{\sqrt{1-4H'^2}} \right),
\end{equation}
will give an algebraic form for the expected probability density, that only depends on the parameter $\sigma_\Delta^2$. Hence, measuring the thermal equilibrium distribution of our pulse outputs and fitting it, allows us to relate the pulse outputs to the parameter $H'$, and further, in the linear regime, to $x_n=H'/\beta$. This fit is plotted together with data in Fig.~\ref{fig:supp_calibration}.

In order to avoid any errors caused by low frequency drifts the thermal histogram points are extracted by taking the difference of two pulses half an oscillation period apart and dividing by two. For one mode, this would give exactly the same histogram as random sampling. Although the fact that we have two modes with differing frequencies produces an error to this procedure, the two frequencies are close enough that the error in $\sigma_{\Delta}^2$ is less than 0.5\% in all cases considered in this paper. (The error is the difference between a factor of $8$ and factor $\left(6-2\cos(r\pi)\right)$. See equation (S21) and compare it to two times equation (S17), taking $\theta=\pi$.)

It should also be emphasized that with our parameters, the expected distribution will always be doubly peaked around $H'=\pm0.5$. Any small uncertainty in parameter $\sigma_\Delta^2$ will not change the position of the peaks. Hence, our calibration is relatively insensitive to any imprecision in $\sigma_\Delta^2$. On the other hand, the vertical scale here is fixed (with proper normalization) and hence the correctness of the height of the center flat part of the histogram will show the accuracy of our $\sigma_\Delta^2$ extraction. As can be seen in Fig.~\ref{fig:supp_calibration} (and this was true generally) the line does lie somewhat above the measured histogram (this error would correspond to $\sim 10\%$ higher $\sigma_\Delta^2$). We note that if this error would be in the $\beta$ parameter, this would make the extracted imprecisions in the main manuscript lower (meaning, better) by $\sim 5\%$.

Finally, we note that we obviously cannot use this method for the off-resonance data we extract the noise floor and single-pulse imprecision from. For that data we first gathered data on-resonance to run the above analysis and then without changing anything except laser frequency run the off-resonance measurement and used the on-resonance data for calibration.

\subsection{Post-selection threshold dependance}

\begin{figure}[h]
\includegraphics[width=0.35\textwidth]{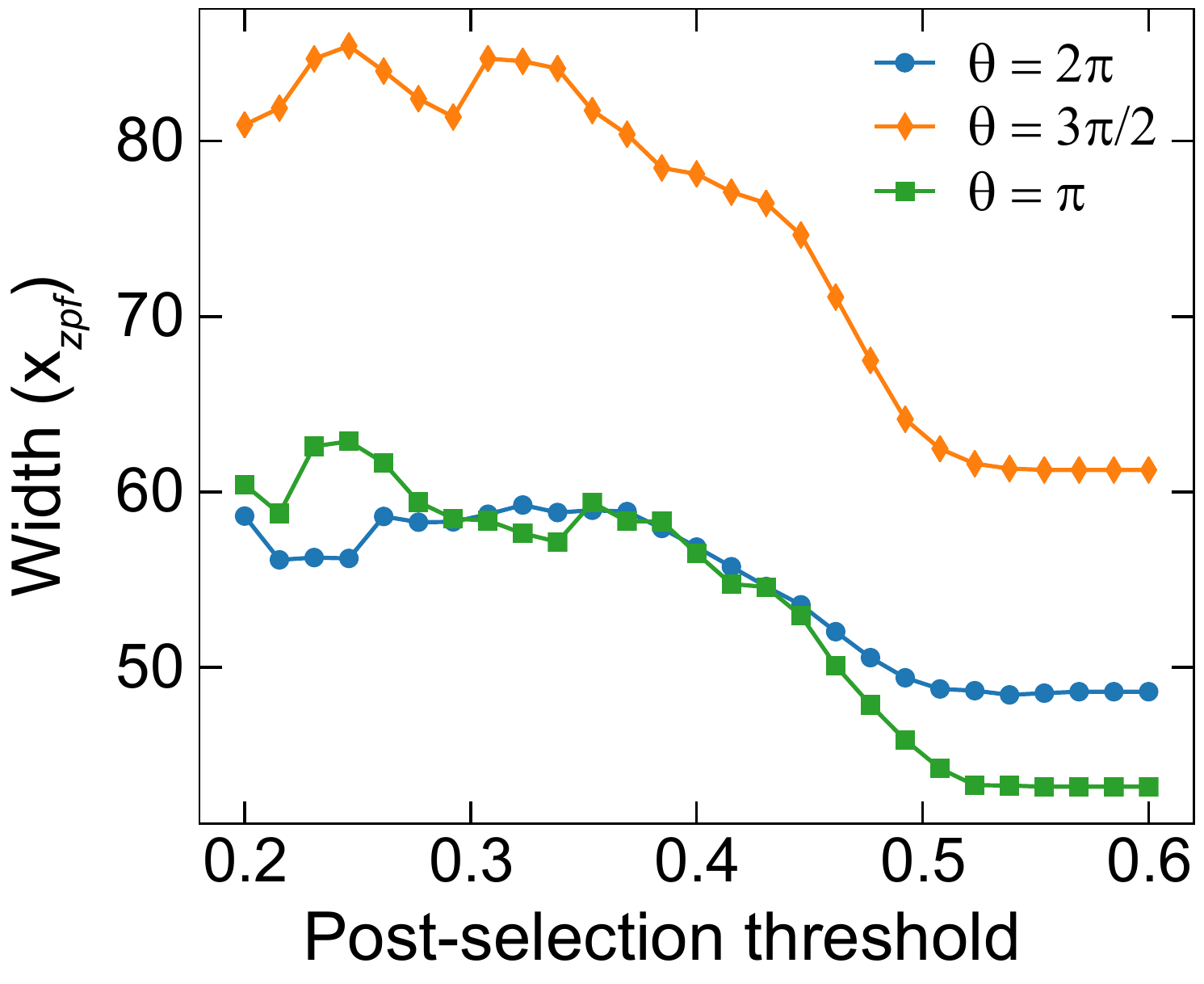}
\caption{Width of different mechanical marginals as a function of the post-selection threshold. The data shown are the three points in Fig.~\ref{fig:three}, corresponding to tomography angles $2\pi$, $3\pi/2$, and $\pi$.
\label{fig:supp_threshold}} 
\end{figure}

As mentioned in the main manuscript, all the data presented has been post-selected so that the instantenous fluctuation amplitude during the measurements is in the linear regime of optomechanical response. In Fig.~\ref{fig:supp_threshold} we show how the conditional state width of data in Fig.~\ref{fig:three}(b) depends on the post-selection point. As expected, the end result is independent of the post-selection point, as long as the point is in the linear regime. If the post-selection point is set higher, the measured width starts to decrease as the measured signal does not correspond linearly to the mechanical displacement.

It is also important to note that the threshold has to be chosen so that there is only a small probability for the higher branch of the transduction function (above $\beta x_n = 1$) to produce the result. For our chosen post-selection threshold the transduction function crosses the threshold first at $0.48\sigma_\textrm{th}$ and then again at $4\sigma_\textrm{th}$. Concretely, this means that we retain roughly one third of the acquired data (per quadrature that is post-selected) and on average only one point in 5000 in that post-selected dataset is from the wrong branch.

\subsection{Homodyne signal and measurement strength}

To derive Eq.~(1) of the main manuscript we consider the interference of two beams each described by the complex field amplitudes: $a_\mathrm{s}$ and $a_\mathrm{lo}$, where the subscripts stand for signal and local oscillator respectively. This happens at the final 50:50 beam splitter of the interferometer and produces the output beams $a_\mathrm{+}=1/\sqrt{2} (a_\mathrm{s} + i a_\mathrm{lo})$ and $a_\mathrm{-}=1/\sqrt{2}(ia_\mathrm{s}+a_\mathrm{lo})$. Each beam is then detected by a photo-detector sensitive to power and the two signals are electronically subtracted. Hence the balanced detector gives as output voltage a signal that is proportional to the difference of the optical powers of the beams $H = \left| a_\mathrm{+} \right|^2 - \left| a_\mathrm{-} \right|^2$. Substituting $a_\mathrm{+}$ and $a_\mathrm{-}$ one obtains
\begin{equation}
H = i (a_\mathrm{s}^\ast a_\mathrm{lo}-a_\mathrm{lo}^\ast a_\mathrm{s}),
\label{eq:BS}
\end{equation}
where the ``$\ast$'' indicates the complex conjugate.

The expression for our signal beam can be obtained from input-output theory. We consider that the signal beam is the output field of the interaction between an incident light beam and the opto-mechanical cavity $a_s = \sqrt{\kappa_\mathrm{out}}a$, where $a$ is the field amplitude inside the cavity. The equation of motion for the field amplitude is
\begin{equation}
\frac{da(t)}{dt}= -i\Delta(t)+\frac{\kappa}{2}a(t) + \sqrt{\kappa_\mathrm{in}} a_\mathrm{in}
\end{equation}
where $\kappa_\mathrm{in}$ is the coupling rate of the incident light with the cavity mode, $\kappa$ is the total cavity energy decay rate and $a_\mathrm{in}$ is the complex amplitude describing the incident beam. An important quantity in this equation is the detuning $\Delta(t) = \Delta_0 + Gx(t)$, where $\Delta_0 = \omega - \omega_\mathrm{c}$ is the static detuning of the laser frequency $\omega$ from the cavity frequency $\omega_\mathrm{c}$. In all measurements in the paper we use $\Delta_0 = 0$. The mechanical system enters the description through the position dependent frequency displacement $Gx(t) = g_0 x_n(t)$. Since in our system the cavity linewidth is large compared to the mechanical frequency $\kappa \gg \omega_\mathrm{m}$, the intracavity field adiabatically follows the evolution of the mechanics. Hence the equation of motion can be adiabatically eliminated leading to 
\begin{equation}
a(t) = \frac{\sqrt{\kappa_\mathrm{in}} a_\mathrm{in}}{-i\Delta(t)+\frac{\kappa}{2}}.
\end{equation}
As the pulse duration in the measurements is also much shorter than the mechanical oscillation period $\tau_P \ll 1/\omega_\mathrm{m}$ we consider the mechanical displacement static during the short measurement interaction and hence $ a(t) \rightarrow a$ as $\Delta(t) = \Delta =  g_0 x_n$ with $x_n$ the static normalized mechanical displacement during the measurement.

Now we can return to Eq.~\ref{eq:BS} and substitute $a_\mathrm{s} = \sqrt{\kappa_\textrm{out}}a$, and $a_\textrm{lo} = |a_\textrm{lo}|e^{i\phi}$, where $\phi$ is the phase difference between the $a_\mathrm{lo}$ and $a_\mathrm{in}$ (hence we can also substitute $a_\mathrm{in}=|a_\mathrm{in}|$). We control this parameter with a mirror attached to a piezo stage. We obtain
\begin{equation}
H= \left| a_\mathrm{lo}\right| \frac{4\sqrt{\kappa_\mathrm{in}\kappa_\mathrm{out}}}{\kappa}\frac{\left| a_\mathrm{in} \right|}{1 + \left(\frac{2\Delta}{\kappa}\right)^2} \left(\cos\phi + \frac{2\Delta}{\kappa}\sin\phi \right),
\end{equation}
which reduces to Eq.~(1) when $\phi = \pi/2$, meaning at the phase sensitive operation point of the homodyne interferometer. Note how the signal is a product of three parts: the amplitude of the local oscillator, the amplitude of the signal beam after interaction with the cavity and the homodyne angle sensitive part. At $\phi=0$ the signal is proportional to only the amplitudes of the two fields, whereas at $\phi= \pi/2$ it becomes linearly sensitive to the detuning of the cavity, meaning in our case the mechanical displacement.

It is instructive to notice that assuming the linear transduction regime in which $4\Delta^2\ll \kappa^2$, and $\phi=\pi/2$ we have
\begin{equation}
H \simeq 4 \left| a_\mathrm{lo}\right| \left| a_\mathrm{in} \right| \eta \beta x_n,
\label{eq:linearhomodyne}
\end{equation}
which we can integrate over the pulse duration
\begin{equation}
\int_0^{\tau_P} H dt \approx \langle H \rangle \tau_P = 4 \langle\left| a_\mathrm{lo}\right|\rangle \sqrt{\tau_P} \langle\left| a_\mathrm{in} \right|\rangle \sqrt{\tau_P} \eta \beta x_n = 4 \sqrt{N_{lo}} \sqrt{N_P} \eta \beta x_n = \sqrt{N_{lo}} \chi x_n,
\end{equation}
using the definition of $\chi$ given in main text and in references \cite{Vanner2011,Hoff2016}. This gives an intuitive feeling for the parameter $\chi$ which is simply the transduction parameter between normalized displacement and the homodyne output. In addition the output is amplified by the strength of the local oscillator as would be expected. As usual, the purpose of this amplification is to bring the signal strength above all electric noise signals. Note that although we have implicitly disregarded some technical variables like the interferometer overlap and the aperture balancing the two detector sides, these can be included in parameter $\kappa_\mathrm{out}$ (and hence $\eta$). Note also that assuming our noisefloor is given by optical shot noise and that the power of local oscillator is much larger than the signal power, the standard deviation of the measurement is given by $\sqrt{N_\mathrm{lo}}$ and hence the signal-to-noise ratio (mean over standard deviation) is simply $\chi x_n$.

\subsection{Conditional state variances}

To illustrate the basic principle of calculating the conditional state variances, consider first the variance of the difference between two pulses separated by angle $\theta = \omega t$, assuming the motion is given by $x(\theta) = X\cos\theta + Y\sin\theta$
\begin{eqnarray}
\Var[x(\theta)-x(0)]_\textrm{1mode}&=&\Var\left[X\cos\theta+Y\sin\theta-X\right]=\Var\left[X(\cos\theta-1)+Y\sin\theta\right] \nonumber \\
&=&\left(\cos\theta-1\right)^2\Var(X)+\sin^2\theta\Var(Y) = (2-2\cos\theta)\Var(Q),
\end{eqnarray}
where we have marked $\Var(Q)=\Var(X)=\Var(Y)$ as the variance of the quadrature amplitudes, which for a thermal state is given by $\Var(Q)=2n_\mathrm{th}x_\mathrm{zpf}^2$. Note that for the difference of two uncorrelated pulses we would expect the variance to be $2\Var(Q)$. 

Any dephasing process can be added to this formula by requiring that the covariance of the quadrature amplitude with itself goes down with time as $\Cov\left[X(t),X(0)\right] = \exp(-\gamma t)\Var[X(0)]$, where $\gamma=\Gamma/2$ for the mechanical damping. Note that this produces the correct limits $\Cov[X(0),X(0)] = \Var[X(0)]$ and $\Cov[X(\infty),X(0)] = 0$. Then using the formula $\Var\left[aX+bX'\right] = a^2\Var[X]+b^2\Var[X']+2ab\Cov[X,X']$ we have
\begin{eqnarray}
\Var\left[X'\cos\theta-X+Y\sin\theta\right]_\textrm{1mode} &=& \cos^2\theta\Var(X')+\Var(X)-2\cos\theta\exp(-\gamma t)\Var(X)+\sin^2\theta\Var(Y) \nonumber \\
&=& \left[2-2\exp(-\gamma t)\cos\theta\right]\Var(Q),
\end{eqnarray}
which produces the expected limits. Note that we assume $\Cov(X,Y)=0$ at all times. Going through similar algebra for the one and two pulse conditional state will produce
\begin{eqnarray}
\Var[x(\theta)-\cos\theta x(0)]_\textrm{1mode}&=& \{1+\cos^2\theta[1-2\exp(-\gamma t)]\}\Var(Q) \\
\Var[x(\theta)-\cos\theta x(0)-\sin\theta x(\pi/2)]_\textrm{1mode}&=& [2-2\exp(-\gamma t)]\Var(Q).
\end{eqnarray}

Now, we want to derive the same formulae in the case of two mechanical modes, $x(\theta) = X_1\cos\theta + Y_1\sin\theta + X_2\cos r\theta + Y_2\sin r\theta$, where $r$ is the ratio of the two mechanical frequencies. For generality we can assume they can have different $n_\mathrm{th}$ and/or $x_\mathrm{zpf}$, and hence $\Var(Q_i)=\Var(X_i)=\Var(Y_i)$, where $i=1,2$. Note that although we do not consider the coupling from the modes to the light field explicitly here, ultimately differing coupling constants would affect the results similarly as differing $x_\mathrm{zpf}$. Going through similar algebra as above will then produce
\begin{eqnarray}
\Var[x(\theta)-x(0)]]_\textrm{2modes} = & & [2-2\exp(-\gamma_1 t)\cos\theta]\Var(Q_1) + [2-2\exp(-\gamma_2 t)\cos r\theta]\Var(Q_2)  \\
\Var[x(\theta)-\cos\theta x(0)]_\textrm{2modes} = & & \{1+\cos^2\theta[1-2\exp(-\gamma_1 t)]\}\Var(Q_1) \nonumber \\
&+& \{1+\cos^2\theta[1-2\frac{\cos r\theta}{\cos\theta}\exp(-\gamma_2 t)]\}\Var(Q_2) \\
\Var[x(\theta)-\cos\theta x(0)-\sin\theta x(\pi/2)]_\textrm{2modes}= & & [(2-2\exp(-\gamma_1 t)]\Var(Q_1) \nonumber \\
&+& \{2-2\exp(-\gamma_2 t) \nonumber \\
&\times &[\cos\theta\cos r\theta+\sin\theta\sin r\theta\sin r\pi/2+\cos r\pi/2\sin\theta(\cos\theta+\cos r\theta)]\}\Var(Q_2) \nonumber \\
\approx & &[(2-2\exp(-\gamma_1 t)]\Var(Q_1) \nonumber \\
&+& \{2-2\exp(-\gamma_2 t)[\cos\theta\cos r\theta+\sin\theta\sin r\theta]\}\Var(Q_2),
\end{eqnarray}
where the last approximation ($\sin r\pi/2=1,\cos r\pi/2=0$), assumes $r$ is sufficiently close to one.

As our pulsing sequence consists actually of more pulses than what is assumed above, the exact analytical formulae are more complicated but the above derivations capture the essential physics. For completeness we quote here the full formula that is plotted with data in Fig.~\ref{fig:three}(b), which is for the pulse sequence given by Eq.~\ref{eq:pulseSequence}
\begin{eqnarray}
\sigma^2
= & & \left[ \frac{1}{8}\left(2 \cos (\theta )+(1-2 \cos (\theta )) \cos (\pi  r)-4 \cos (\theta  r)+(2 \sin (\theta )+1) \cos \left(\frac{3 \pi  r}{2}\right)+(1-2 \sin (\theta )) \cos \left(\frac{5 \pi  r}{2}\right)+1\right)^2 \right. \nonumber \\
&+& \left. \frac{1}{8} \left((2 \sin (\theta )+1) \sin \left(\frac{3 \pi  r}{2}\right)+(1-2 \sin (\theta )) \sin \left(\frac{5 \pi  r}{2}\right)+4 \sin (\theta  r)+(1-2 \cos (\theta )) \sin (\pi  r)\right)^2\right]\Var(Q_2)
\end{eqnarray}
This is without thermal dephasing (meaning $\gamma_i=0$), which means there is no contribution from $\Var(Q_1)$.

\end{document}